\documentclass[aps,twocolumn,groupedaddress,showpacs]{revtex4}
\usepackage{graphics}
\begin{document}
\newcommand{\beq}{\begin{equation}}
\newcommand{\eeq}{\end{equation}}
\bibliographystyle{apsrev}

\title{Popper's experiment, Copenhagen Interpretation and Nonlocality}

\author{Tabish Qureshi}
\email[Email: ]{tabish@jamia.homelinux.net}
\affiliation{Department of Physics, Jamia Millia Islamia, New Delhi-110025,
INDIA.}


\begin{abstract}

A thought experiment, proposed by Karl Popper, which has been experimentally
realized recently, is critically examined. A basic flaw in Popper's
argument which has also been prevailing in subsequent debates, is
pointed out. It is shown that Popper's experiment can be understood
easily within the Copenhagen interpretation of quantum mechanics. An
alternate experiment, based on discrete variables, is proposed, which
constitutes Popper's test in a clearer way. It refutes the argument of
absence of nonlocality in quantum mechanics.

\end{abstract}


\maketitle

\section{Introduction}

Despite the tremendous success of quantum mechanics, since its inception,
it is one theory which has been a subject of constant debate regarding its
interpretation. One might venture to say that it is the most successful,
but the least understood theory. ``{\em I look upon quantum mechanics with
admiration and suspicion}", wrote Albert Einstein in a letter to 
P. Ehrenfest.  The so called EPR argument, first posed by Einstein,
Podolsky and Rosen (EPR) in 1930s\cite{epr}, sums up the
discomfort with the picture of the physical world that the quantum
theory suggests.  The EPR thought experiment, later reformulated in
terms of spin-1/2 particles by Bohm, has been interpreted as showing,
what Einstein called, ``spooky action at a distance". Bohr\cite{bohr}
tried to counter this argument saying that various definitions are tied
to the experimental setup used and cannot be decoupled from it. The very
act of measurement can influence the physical reality.  The essence of
Bohr's argument constitutes the {\em Copenhagen interpretation} of quantum
mechanics where the wavefunction of a multi-particle system is regarded
as one, and disturbing any part of it, can disturb the whole system.
In this view, a measurement on one particle can have a non-local influence
on a spatially separated particle, even in the absence of any physical
interaction. In the quantum mechanics lore, this has come to be known as
{\em nonlocality}.

Twentieth century philosopher of science, Karl Popper believed
that quantum formalism could be interpreted realistically. He
proposed an experiment to demonstrate that a particle could have a
precise position and momentum at the same time.  For some reason,
Popper's thought experiment did not attract much attention from the
physics community, but there has been a recent resurgence of debate on
it\cite{krips,sudbury,collet,storey,redhead,plaga,short,nha,peres,hunter}.
New interest has also been
generated by its actual realization by Kim and Shih\cite{shih}, and also
by claims that it proves the absence of quantum nonlocality\cite{unni}.

In this paper, we critically analyze Popper's thought experiment and its
realization, and point out the flaws in the argument. We also
propose, what we think is, a discrete version  of Popper's experiment.
We believe, this thought experiment captures the essence of ``Popper's test",
and serves to clarify the issues regarding the Copenhagen interpretation and
quantum nonlocality.

\section{Popper's thought experiment}

Let us start by describing the thought experiment Popper proposed. Basically
it consists of a source $S$ which can generate pairs of particles traveling
to the left and to the right, which are entangled in the momentum space.
This is to say that momentum along the y-direction of the two particles is
entangled in such a way, so as to conserve the initial momentum at the source,
which is zero. There are two slits, one each in the paths of the two particles.
Behind the slits are sitting arrays of detectors which can detect the particles
after they pass through the slits (see Fig. 1a).
\begin{figure}
\resizebox{3.5in}{!}{\includegraphics{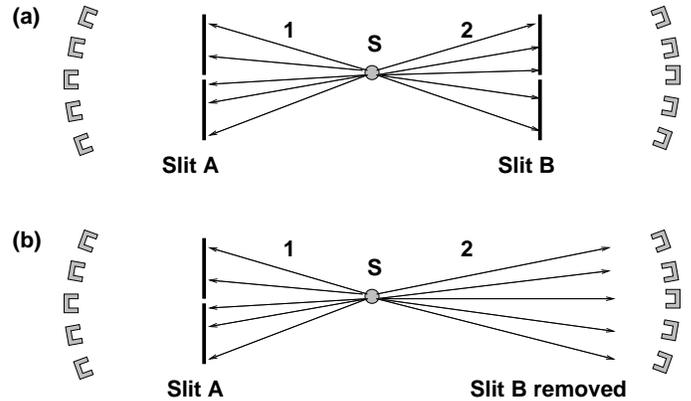}}
\caption{Schematic diagram of Popper's thought experiment. (a) With both the
slits, the particles are expected to show scatter in momentum. (b) Removing
slit B, Popper believed, one could test the Copenhagen interpretation. }
\end{figure}

Being entangled in the momentum space implies that in the absence of the two
slits, if particle on the left is measured to have a momentum $p$, the particle
on the right will necessarily be found to have a momentum $-p$. One can
imagine a state similar to the original EPR state \cite{epr}, 
$\psi(y_1,y_2) = \int_{-\infty}^{\infty}e^{ipy_1/\hbar} e^{-ipy_2/\hbar}dp$.
As one can see, this state also implies that if particle on the left is
detected at a distance $y$ from the horizontal line, the particle on the right
will necessarily be found at the same distance $y$ from the horizontal line.
In the
presence of the slits, Popper argued, when the particles pass through the slits,
they experience a large uncertainty in momentum. This results in
a larger spread in the momentum, which will be show up as particle being
detected even at positions which are away from the line connecting the source
and the slit. This spread, because of a real slit is expected.
A tacit assumption in Popper's setup is that the initial spread in momentum
of the two particles is not very large.

Popper then suggests that slit B be removed. In this situation, Popper argues
that when particle 1 passes through slit A, it is localized in space, to
within the width of the slit. If one believes in the Copenhagen interpretation
of quantum mechanics, then one would think that when particle 1 is 
localized in
space, particle 2 should also get localized in space. In fact, if we do this
experiment without the slits, the correlation in the detected positions of 
particles 1 and 2, implies just this. This is the collapse postulate of quantum
mechanics, for which no mechanism is given. And an act of measurement on
particle 1, seems to have a spooky action on the particle 2. But Popper had
something else in mind. He was not convinced by the correlation in the 
detected positions of the particles. He argued that if particle 2
actually experiences a localization in position, its subsequent evolution
should show a larger spread
in momentum. To be precise, just as much as there was when the real slit B was
present. Popper had his own argument to suggest that if such an experiment is
actually performed, no extra spread in momentum will be observed. This, he said,
showed that Copenhagen interpretation doesn't work. 

\section{Is nonlocality absent?}

Based on Popper's thought experiment, an argument has been put forward
by Unnikrishnan\cite{unni} which claims that there is no nonlocality in
quantum mechanics. The argument is as follows. If there were an actual
reduction of the state when the particle 1 went through slit A, particle
2 would get localized in a narrow region of space, and in the subsequent
evolution, experience a greater spread in momentum. If no extra spread
in the momentum of particle 2 is observed, it implies that there is no
nonlocal effect of the measurement of particle 1 on particle 2. The tacit
assumption here is that the correlation observed in the detected positions
of particles 1 and 2, in the absence of the slits, could be explained in
some other way, without invoking a nonlocal state reduction.

\section{Realization of Popper's experiment}

Popper's thought experiment was recently realized by Kim and Shih using
an entangled two-photon source \cite{shih}. They used a modified geometry as 
demanded by the experimental arrangement. When both the slits A and B are
present, they observed a significant spread in the momentum, seen as
a scatter in the detected positions of the photons. When slit B is removed,
particle 1 shows a spread in momentum, but particle 2 doesn't show any spread.
This is in agreement with what Popper had predicted.
Infact, particle 2 is observed to lie in a region which is much 
narrower than the initial spread of the beam.
But the question is, does it indicate that Copenhagen interpretation is
flawed, or that nonlocality is absent? It should be mentioned here that
there have been objections against Kim and Shih's experiment, pointing out
that due to the finite size of the source, the localization of the second
particle is not perfect \cite{short}. We will come back to this point at
the end of the discussion.

\section{Analysis of the experiment}

One problem with the original proposal is that the source is assumed to
have a sharp momentum value, which would imply that the position of the
source is uncertain by an amount dictated by the uncertainty principle.
This was pointed out by Collet and Loudon \cite{collet} who concluded that
due to this uncertainty, the experiment would not be able to test quantum
mechanics. Similar arguments have been put forward by others, like Redhead
\cite{redhead}. This is a valid criticism, but it turns out that precisely
this kind of setup is not crucial for Popper's experiment. Now spontaneous
parametric down conversion (SPDC) in nonlinear optical crystals can yield
particles which are entangled in precisely the way required by Popper's
experiment \cite{shih}. The measured positions of such particles are
correlated with nearly unit probabilty \cite{strekalov}. Although the
use of finite size source has been a source of controversy\cite{short},
we address the question that {\em if such particle pairs can be produced
(maybe with a more extended source, or by some other means), what will
be its consequence for Popper's experiment}.  In order to follow the
language of Popper's original proposal, we will continue to use a point
source in the subsequent discssion, but it should be kept in mind that
the argument can be easily translated to the case of SPDC particle pairs
coming from an extended source.

Let us try to analyze the experiment carefully and see what result one
would expect within the Copenhagen interpretation of quantum mechanics.
Popper argued that according to Copenhagen interpretation, when particle 1 
passes through slit A, the wavefunction should get reduced to something which
is localized within the width of the slit. But let us ask the question,
when do we acquire the knowledge that the particle has passed through the slit.
The answer is, {\em not until particle 1 has been detected by one of the detectors.}
To reinforce this point, let us assume that we had put an array of detectors
next to slit A. In that situation, particle would either get
detected by one of the detectors near slit A, or pass through the slit and
get detected by the detectors behind the slit. So, we can have knowledge
that the particle passed through the slit or not if, either it is detected by
the detectors behind the slit, or the ones next to it. Now, in Popper's
experiment, we are only interested in the particles which have passed
through slit A, and not in the ones that could not pass through, and
are lost somewhere else. In this situation, we can only know that  particle 1
passed through
the slit, when one of the detectors behind the slit detects it. {\em This is
the point at which one has to invoke a measurement, and a reduction of the
state,} and not at the point when the particle reaches the slit. The detector
behind slit A {\em causes} a reduction of the state.

If one agrees with this argument, then the analysis of both Kim and Shih
\cite{shih} and Short \cite{short}, which talk of ``localization" of the
wavefunction at the slit, are questionable. 

This is a fundamental flaw in Popper's argument, which leads him to
believe that Copenhagen interpretation doesn't work. Let us now find out
what happens on the right, that is, what does particle 2 do in this situation.
Remember that particles 1 and 2 are entangled in momentum states.
If particle
1 is found to have momentum $p'$, then particle 2 will necessarily be found
to have a momentum $-p'$.
EPR like states have a property that detected positions of particles
are correlated.
Although, the momentum spread is assumed to be not very large, we will assume
that such a correlation is possible, and will explore its consequence.
One can see that with this kind of entanglement, the angular
directions of the two particles get correlated, which can also be
experimentally verified.

Now for particle 1, only directions which lie approximately within a small
angle (see Figure 2a) will allow it to pass through slit A. Other parts of 
the wavefunction will be blocked by the slit wall. This is a direct
consequence of the entanglement in momentum, as detecting particle 2 on the
right hand side, gives us the {\em ``which path"} information about particle 1.
For the very same reason, if one were to carry out a double-slit interference
experiment on particle 1, no interference would be seen. 

Now, the interesting thing is that, because of entanglement, this part
of the wavefunction also describes particle 2. If
particle 1 passes through slit A, whatever evolution it goes through 
subsequently, does not affect the entangled part of the particle 2
wavefunction. Particle 1 may experience a spread in the wavefunction because of
slit A, but the entanglement with particle 2 remains intact. As a result,
{\em particle 2 continues to evolve as it would have, if slit A were absent.}
When particle 1 is detected behind slit A, particle 2 will be detected at
a position which lies within a narrow angle which corresponds to particle 1
passing through the region of slit A, {\em but in the absence of the slit}
(see Fig 2b).
This is so, because the part of the particle 1 wavefunction which passes
through slit A, is entangled to only that particular part of the 
particle 2 wavefunction. This all happens with a certain probability.
There is also a probability  that particle 1 doesn't enter slit A. In that
case, particle 2 will be detected at other positions.
As mentioned before, Popper's experiment doesn't consider this case, and
we will not discuss it here.

\begin{figure}
\resizebox{3.5in}{!}{\includegraphics{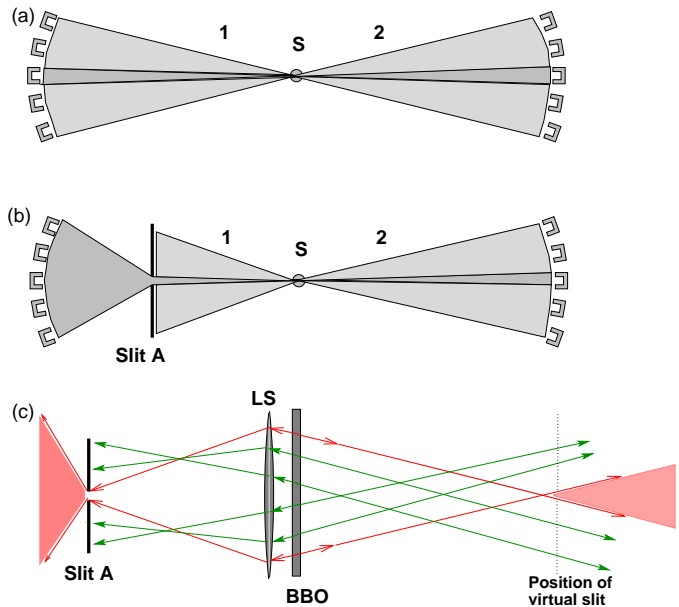}}
\caption{Schematic diagram to understand what happen's in Popper's thought
experiment. Due to entanglement in momentum, each direction of particle
1 is correlated to a direction of particle 2. (a) In the absence of slit A,
there is a region which corresponds to particle 1 passing within a region
where slit A could be present. (b) In the presence of slit A, particle
1 is scattered, but particle 2 travels unaffectd. (c) With a broad source,
like a BBO crystal, some directions (indicated in red)
contribute to the particles passing through the region of the slits. Other
directions (indicated in green) contribute to particle going to regions
outside the slits.}
\end{figure}

This argument can be easily applied to the case of an extended source,
as used in Kim and Shih's experiment. Figure 2(c) gives a schematic
representation of what happens in such the case where a combination of
a BBO crystal and a converging lens is used. In the language of photons,
only some $k$ values in different regions of the BBO crystal will contribute
to the photons passing through the regions of the slits. These directions,
indicated by red lines, for the two particles are correlated. Of course
there are lot of other $k$ values, indicated in green, which correspond
to particles not passing through the slits. The end result is that the
parts of the wavefunctions of the particles indicated in pink, on both sides,
are correlated with each other.

So, the conclusion of the preceding discussion is that {\em if particle
1 is detected by any detector behind slit A, particle 2 will be found to
have a position which lies within a narrow angle which corresponds to
particle 1 passing through the region of slit A, in the absence of the
slit.} This is the conclusion of the Copenhagen interpretation. As one
can see, this is in sharp contrast to what Popper had concluded regarding
the Copenhagen interpretation.  This also appears to be in agreement with
the experimental result of Kim and Shih\cite{shih}. However, if the finite
size of the source indeed led to unsatisfactory correlation between the
photon pairs, it might be an interesting exercise to repeat the experiment
with a better source. In that case, Short's work will predict a larger
momentum spread in the second particle. On the other hand, we predict
that a better correlation would not lead to an increase in the momentum
spread of the second particle, as argued in preceding discussion.

\section{A discrete version of Popper's test}

One reason for which Popper's experiment has been criticized is that it
uses continuous variables, and it is not clear at what stage is invoking the
uncertainty principle justified. As we saw in the preceding discussion, Popper's
experiment fails to achieve what Popper aimed at. The essence of Popper's
argument, at least as far as nonlocality and the Copenhagen interpretation
are concerned, is not based on the
precise variables he chose to study, namely position and momentum. Any
two variables which do not commute with each other should serve the
purpose, as localizing one would lead to spread in the other. This point
has also been emphasized by Unnikirshnan\cite{unni}.
In the
following, we present a discrete model which, we believe, captures the
essence of Popper's  test.

\subsection{The model}

Consider two spin-1 particles $A$ and $B$, emitted from a source $S$
such that $A$ travels along negative $y$ direction, and $B$ travels along
positive $y$ direction. The particles start from a spin state which is
entangled in such a way that if z-component of the $A$ spin is found to
have value $+1$, the z-component of $B$ will necessarily have value
$-1$. The initial spin state of the combined system can be written as
\begin{equation}
|\psi\rangle = \alpha |A_z;+1\rangle|B_z;-1\rangle
	       + \beta |A_z;0\rangle|B_z;0\rangle
	       + \alpha |A_z;-1\rangle|B_z;+1\rangle
\end{equation}
where $|A_z;m\rangle$ and $|B_z;m\rangle$ represent the eigenstates of
the z-component of the spins $A$ and $B$ respectively, with eigenvalue $m$.
Also, the state $|\psi\rangle$ is normalized, so that $2\alpha^2+\beta^2=1$.
Here, the z-components of the spins can be thought as playing the
role of momenta in the $y$ direction of the two particles in Popper's
experiment. In that case, the x-component of the spin here can play the role
of position of the two particles along $y$ axis, in Popper's experiment.
The two components of the spin do not commute with each other, so localizing
one in its eigenvalues, will necessarily cause a spread in the eigenvalues
of the other. Thus, this spin system is completely analogous, in spirit, to 
the system of entangled particles, considered by Popper. 

Next, we have to have a mechanism which is equivalent to localizing
the particle 1, in Popper's experiment, in space (what he wanted to
achieve by putting a slit). To achieve this, we put a Stern-Gerlach
field in the path of particle $A$, pointing along the $x$ axis, but
inhomogeneous along the (say) z-axis. This will split the particle $A$
into a superposition of three wave packets, spatially separated in the
$z$ direction, entangled with the three spin states $|A_x;+1\rangle$,
$|A_x;0\rangle$ and $|A_x;-1\rangle$. Then we put a detector $D1$ in the path
of this particle such that, it detects the central wave packet and localizes
the x-component of spin $A$ to the state $|A_x;0\rangle$. This achieves,
what slit A was supposed to achieve in Popper's experiment, but actually
never did, namely localizing the particle in position.
\begin{figure}
\resizebox{3.5in}{!}{\includegraphics{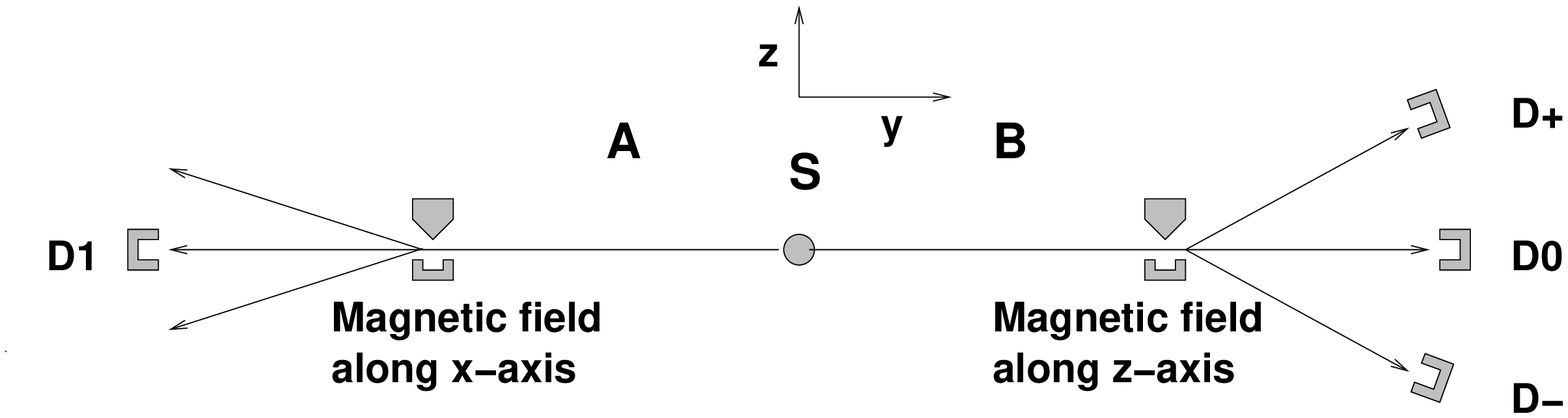}}
\caption{Schematic diagram of a discrete version of Popper's experiment.
Detector $D1$ detects particle A, and particle B is detected by
detectors $D+$, $D0$ and $D-$.}
\end{figure}

On the other side of the source, we can have a Stern-Gerlach field, in the
path of particle $B$, pointing along the z-direction. This will split
particle $B$ into a superposition of three wave-packets, entangled with
the three spin states $|B_z;+1\rangle$, $|B_z;0\rangle$ and $|B_z;-1\rangle$.
We have three detectors, $D+$, $D0$ and $D-$, to detect one component each
of the $z$-component of spin $B$.

\subsection{What do we expect?}

Now, the $z$-components of spins $A$ and $B$ are entangled. So,
it is indisputable that if one finds $A$ in $|A_z;+1\rangle$ state,
$B$ would be found in $|B_z;-1\rangle$ state, and if one finds $A$ in
$|A_z;-1\rangle$ state, $B$ would be found in $|B_z;+1\rangle$ state,
and so on. Also, one can easily verify that if one measures the $x$
component of spin $A$ and finds it in the state $|A_x;0\rangle$, one would
find the $x$-component of spin $B$ in the state $|B_x;0\rangle$. But,
as operators $B_x$ and $B_z$ do not commute, if one finds spin $A$ in
the state $|A_x;0\rangle$, there should be a spread in the eigenstates
of $B_z$.  In Popper's experiment, this would be equivalent to saying,
that if particle 1 is localized in {\em position}, there should be a
spread seen in the {\em momentum} of particle 2. This is what the Copenhagen
interpretation predicts. At this stage, the
equivalence of this experiment with Popper's experiment is complete.

In addition, if one applies Unnikrishnan's argument\cite{unni} to the 
present model, detecting particle $A$ in the detector $D1$ leading to
observation of a spread in the counts of particle $B$ in the three detectors,
amounts to a nonlocal {\em action at a distance}.

\subsection{``Doing" the thought experiment}

Let us now carry out this thought experiment and see what we get. 
To start with, we first remove the detector and the Stern-Gerlach field
from the path of particle $A$. We start from a spin state
$|\psi\rangle$ where $\beta = \sqrt{0.9}$ and $\alpha = \sqrt{0.05}$, which
has the following form:
\begin{eqnarray}
|\psi\rangle &=& \sqrt{0.05} |A_z;+1\rangle|B_z;-1\rangle
	       + \sqrt{0.9} |A_z;0\rangle|B_z;0\rangle \nonumber\\
	       && + \sqrt{0.05} |A_z;-1\rangle|B_z;+1\rangle \label{teststate}
\end{eqnarray}
It is trivial to see that the three detectors on the right will click
in the following manner. The detector $D0$ will show 90 percent
counts and the other two will have 5 percent each (see Fig. 4a).

Next we put the Stern-Gerlach field and the detector in the path of
particle $A$. As in Popper's experiment, we have to do coincident count
between the detector on the left, and the detectors on the right.
As we are measuring the $x$-component of the spin $A$
on the left, it would be natural to write the state (\ref{teststate})
in terms of the eigenstates $|A_x;m\rangle$. In this form, the state
$|\psi\rangle$ looks like
\begin{eqnarray}
|\psi\rangle &=& |A_x;+1\rangle({\sqrt{0.05}\over 2}|B_z;+1\rangle
                + {\sqrt{0.9}\over \sqrt{2}}|B_z;0\rangle  \nonumber\\
             && + {\sqrt{0.05}\over 2}|B_z;-1\rangle) \nonumber\\
             && - \sqrt{0.05}|A_x;0\rangle({1\over \sqrt{2}}|B_z;+1\rangle
                - {1\over \sqrt{2}}|B_z;-1\rangle) \nonumber\\
             && + |A_x;-1\rangle({\sqrt{0.05}\over 2}|B_z;+1\rangle
                - {\sqrt{0.9}\over \sqrt{2}}|B_z;0\rangle  \nonumber\\
             && + {\sqrt{0.05}\over 2}|B_z;-1\rangle)\nonumber\\ \label{fstate}
\end{eqnarray}
It is clear from (\ref{fstate}), that in a coincident count between the
detector on the left and the detectors on the right, spin $A$ is found
in state $|A_x;0\rangle$ by choice, and spin $B$ ends up in the state
${1\over \sqrt{2}}(|B_z;+1\rangle - |B_z;-1\rangle)$. This means that the
detectors on the right will have 50 percent count each in the detectors $D+$
and $D-$, and no count in the detector $D0$! (see Fig. 4b) 
To start with, the z-component of spin $B$ was predominantly localized in
the state $|B_z;0\rangle$, as seen in the experiment without the detector
and the field for particle $A$. Localizing the spin $A$ in the state
$|A_x;0\rangle$, results in a large scatter in the $z$-component of spin $B$.

\begin{figure}
\center{
\resizebox{2.6in}{!}{\includegraphics{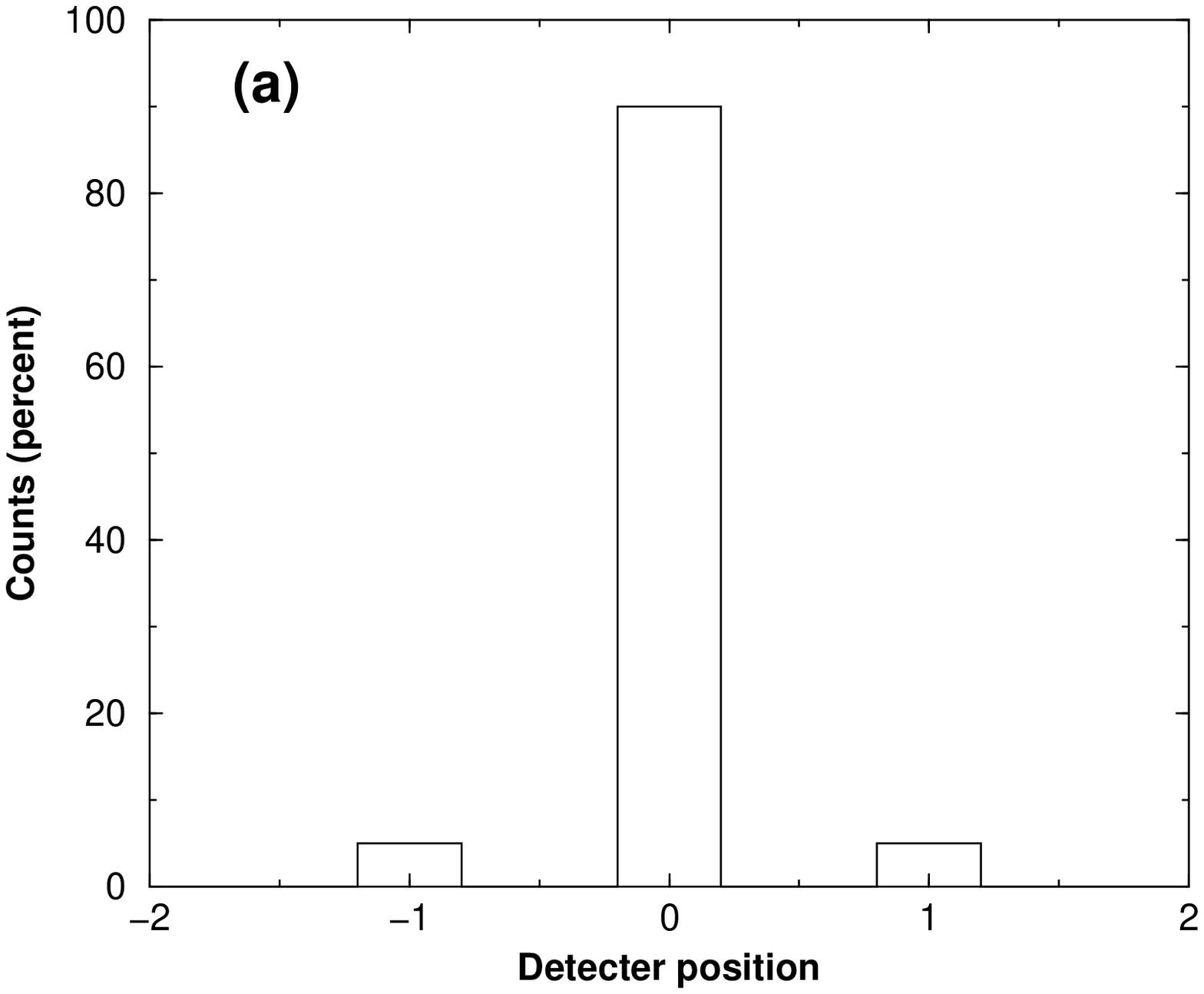}}
\resizebox{2.6in}{!}{\includegraphics{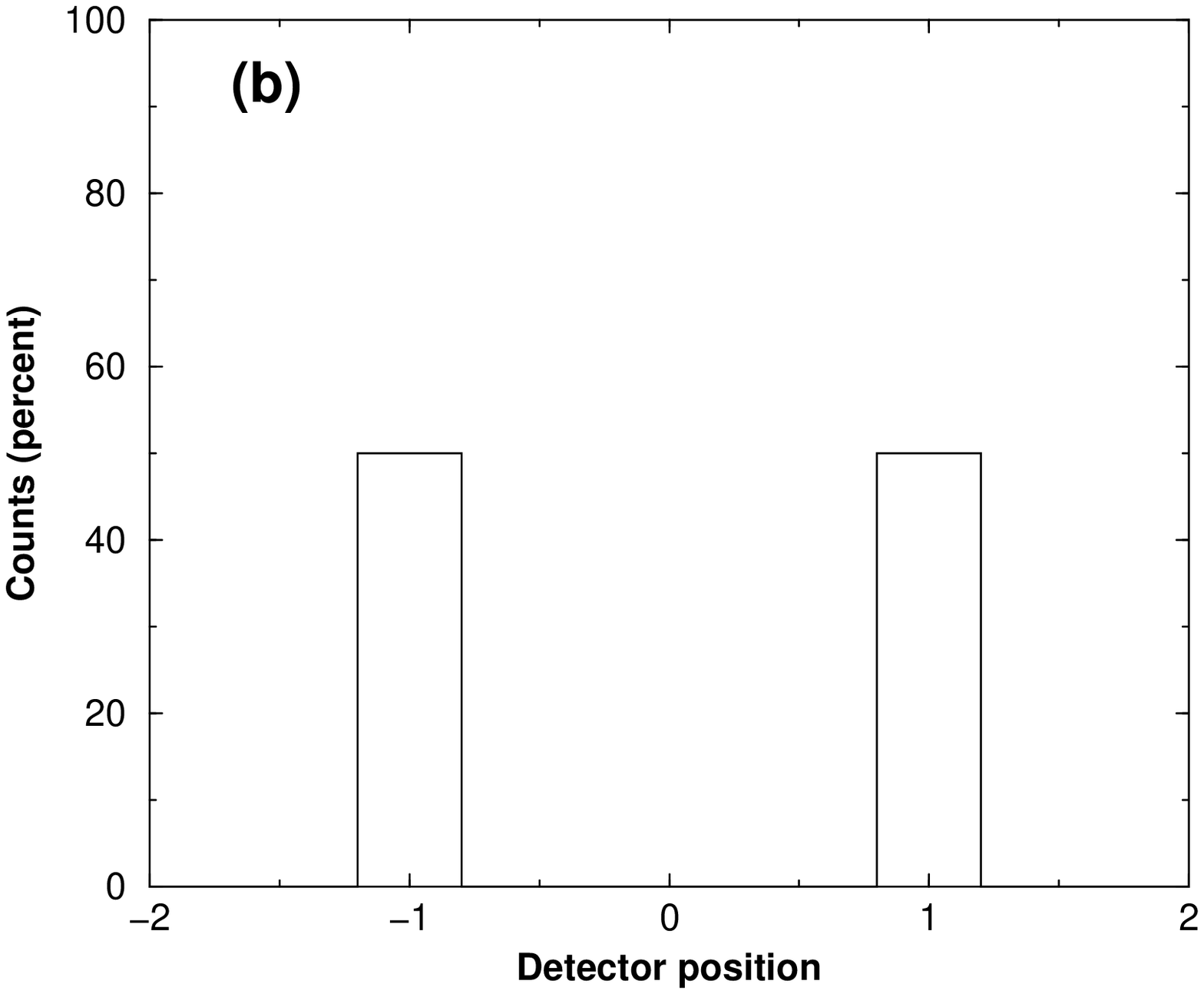}}}
\caption{Results that the detectors $D+$, $D0$ and $D-$ are expected to show
(a) without the detector
$D1$ and the Stern-Gerlach field in the path of particle A, and (b)
with coincident counting with the detector $D1$. Here, detector positions
$-1, 0, +1$ correspond to the detectors $D-$, $D0$ and $D+$ respectively.}
\end{figure}

In Popper's experiment, this will be equivalent to saying that localizing
particle 1 in space, leads to a scatter in the momentum of particle 2.
Thus we reach the same conclusion that Popper said, Copenhagen interpretation
would lead to. But the difference here is that, looking at (\ref{fstate})
nobody would say that in actually doing this experiment, one would not
see the result obtained here. This comes out just from the mathematics of
quantum mechanics, without any interpretational difficulties, as in Popper's
original experiment.

\section{Afterthoughts}

Now that we are through with discussing both the models, let us find out
why this model gives us something which Popper's thought experiment was
unable to. As discussed before, the main flaw in Popper's argument was in
assuming that the Copenhagen interpretation implies that the slit causes
a reduction of the state immediately. The discrete model introduced here,
relies on an actual detection of the particle, which causes a reduction of
the state. This results in a spread in the $z$-component of spin $B$. So,
our conclusion is that the Copenhagen interpretation passes Popper's
test, only that, in our view, this discrete model is the right way to
implement it.

Regarding the issue of nonlocality, we can pose the question whether this
discrete model really shows ``spooky action at a distance". On the face of
it, Fig. 4 seems to suggest that. Without the detector $D1$, the detectors
$D+$, $D0$ and $D-$ show counts primarily concentrated at $D0$, and with
coincident counting with $D1$, the detector $D0$ shows no counts and $D+$
and $D-$ show 50 percent counts each. Knowledge of particle A being at
$D1$, seems to increase the dispersion in particle $B$. 
But a careful look reveals that
the overall count distribution for particle $B$ is still the same as
before. We are only choosing those counts that are coincident with $D1$,
and are throwing away the rest. So, a defender of locality can argue that
we are not {\em affecting} particle $B$ by doing something
to particle $A$. We just see a correlation in various events on the left
and the right. In a sense, that is true - the nonlocality that we observe
in this discrete model, is only at the level of observing correlations
between the spatially separated particles, as is the case with any
experiment done till date. Some people like to believe that just the
presence of correlations does not prove nonlocality - correlations could
be explained in some other way, like by introducing hidden variables, and that
it only shows the ``incompleteness" of the quantum formalism.  This work
doesn't throw any new light on that debate.  However, from Popper's
argument, and that of several others, it appears that they would have
accepted nonlocality had they found a positive result of Popper's test.
In this light, the result of the present work can be considered a
signature of nonlocality.

So, we have concluded that because the slit does not reduce the wavefunction,
we should not expect any increase in the momentum spread of the second
particle in Popper's experiment. An immediate thought that comes to mind is
that if slit A were replaced by a real narrow detector, do we expect to
see an increase in the momentum spread of particle 2, and if so, what would
that indicate?
By analysing the discrete model, we have gained some clarity regarding Popper's
experiment. In the discrete model, we saw that the two peaks which appear
in the conicidence counting, were already present in the initial state
as the two smaller peaks which were overshadowed by the bigger central peak.
So, the ``spread" we see in the coincident counts, was already present in
the original state. In Popper's experiment, if we expect to see a spread
in momentum of particle 2 in the coincident counts with particle 1 behind
slit A, it has to be already present in the wavefunction. But what about
Popper's orginal argument that if particle 2 gets localized in a narrow
region of space, it should have an increased momentum spread? The catch
is that particle 2 can be localized in a narrow region of space {\em only
if} the wavefunction is of the form
$\psi(y_1,y_2) = \int_{-\infty}^{\infty}e^{ipy_1/\hbar} e^{-ipy_2/\hbar}dp$,
which means that momentum spread is infinite! If the momentum spread is
already infinite, particle 2 cannot show any extra momentum spread. If,
on the other hand, the momentum spread is finite, it will not be possible
to precisely localize particle 2 in a narrow region of space, and
subsequently one should not expect large momentum spread. Short's
criticism of Kim and Shih's experiment correctly points out that due to
the finite size of the source, the localization of particle 2 will not be precise, but
doesn't say what happens if the source is improved \cite{short}. Others,
including Sudbury \cite{sudbury} have recognized the problem with infinite
mometum spread being already present in the initial state.
So, what is fundamentally wrong
in Popper's proposal is the assmuption that according to the
Copenhagen interpretation, using entangled particles with a {\em finite}
momentum spread, particle 2 can be localized precisely in position. With this 
knowledge one can be sure that {\em in Popper's experiment, no extra spread in
momentum, which is not already present in the initial state, can ever be seen}.
This is independent of the way in which Popper's experiment is realized.

With this understanding, we believe, the controversy regarding Popper's
experiment is fully resolved.  

\section{acknowledgements}

The main impetus for this work came from the stimulating discussions
during the meeting ``{\em The Quantum World: Frontiers, Foundations and
Philosophy}" organized by the {\em Centre for Philosophy and Foundations of
Science}, in New Delhi in December 2002.


\begin{thebibliography}{0}
\bibitem{epr} A. Einstein, B. Podolsky and N. Rosen, Phys. Rev. {\bf 47}, 
777 (1935).

\bibitem{bohr} N. Bohr, Nature {\bf 136}, 65 (1935).

\bibitem{popper} K.R. Popper, in {\em Open Questions in Quantum Physics},
G. Tarozzi and A. van der Merwe, eds. (Reidel, Dordrecht, 1985);\\
K.R. Popper, {\em Quantum Theory and the Schism in Physics} (London:Hutchinson)
pp. 27-29 (1982).

\bibitem{krips} H. Krips, {\em Brit. J. Phil. Sci.} 253 (1984).

\bibitem{sudbury} A. Sudbury, {\em Phil. Sci.} {\bf 52}, 470 (1985).

\bibitem{collet} M.J. Collet and R. Loudon, {\em Nature} {\bf 326}, 671
(1987).

\bibitem{storey} P. Storey, M.J. Collet and D.F. Walls, Phys. Rev. Lett.
{\bf 68}, 472 (1992).

\bibitem{redhead} M. Redhead in {\em Karl Popper: Philosophy and Problems},
ed. A. O'Hear (Cambridge) (1996).

\bibitem{plaga} R. Plaga, {\em Found. Phys. Lett.} {\bf 13}, 461 (2000).

\bibitem{short}  A. J. Short, {\em Found. Phys. Lett.} {\bf 14(3)}, 275 (2001).

\bibitem{nha} H. Nha, J-H. Lee, J-S. Chang and K. An, Phys. Rev. A {\bf 65},
033827 (2002).

\bibitem{peres} A. Peres, {\em Studies in History and Philosophy of Science}
{\bf 33} (2002) 23.

\bibitem{hunter} G. Hunter, {\em AIP Conference Proc.} {\bf 646(1)}, 243-248
(2002).

\bibitem{shih} Y-H Kim and Y. Shih, Found. Phys. {\bf 29}, 1849 (1999).

\bibitem{unni} C.S. Unnikrishnan, Found. Phys. Lett. {\bf 15}, 
1 (2002).

\bibitem{strekalov} T.B. Pittman, Y.H. Shih, D.V. Strekalov, A.V. Sergienko,
Phys. Rev. A, {\bf 52}, R3429 (1995).

\end{thebibliography}
\end{document}